\documentstyle[pre,aps,multicol]{revtex}

\def\tou#1{{\lower1.2ex\hbox{$\longrightarrow$}\atop
        {\lower-.7ex\hbox{$\scriptscriptstyle #1 $}}}}
\def\lsim{{\lower1.2ex\hbox{$<$}\atop
        {\lower-.7ex\hbox{$\sim$}}}}
\def\gsim{{\lower1.2ex\hbox{$>$}\atop
        {\lower-.7ex\hbox{$\sim$}}}}

\def\be{\begin{equation}}
\def\ee{\end{equation}}
\def\bea{\begin{eqnarray}}
\def\eea{\end{eqnarray}}

\tighten
\input epsf

\begin{document} 
\draft
\raggedbottom

\title{\Large \bf  
Short-time Critical Dynamics of the \\
3-Dimensional Ising Model\thanks{
Work supported in part by the Deutsche 
Forschungsgemeinschaft;
DFG~Schu 95/9-1 and SFB~418}}

\author{\bf A. Jaster, J. Mainville, L. Sch\"ulke and B. 
Zheng$^{\S}$}

\address{Universit\"at -- GH Siegen, D -- 57068 Siegen, 
Germany}

\address{$^{\S}$Universit\"at Halle, D -- 06099 Halle, 
Germany}

\maketitle

\begin{abstract}
Comprehensive Monte Carlo simulations
of the short-time dynamic behaviour
are reported 
for the three-dimensional Ising model at criticality. 
Besides the exponent $\theta$ of the critical initial increase and
the dynamic exponent $z$, the static critical exponents $\nu$ and $\beta$
as well as the critical temperature 
are determined from the power-law scaling behaviour
of observables
at the beginning of the time evolution.
States of very high temperature as well as of zero temperature
are used as initial states for the simulations.
\end{abstract}


\begin{multicols}{2}\narrowtext

\section {Introduction}

Critical properties of many magnetic materials 
can be described  by a simple Ising model,
\be
H=K\;\sum_{<i,j>}\; S_i S_j\;,
\label{ham}
\ee
where $S_i=\pm1$ represents the spin of site $i$,
and the sum extends over nearest neigbours only.
The factor $1/k_BT$ is included in
the coupling constant $K$.
The Ising model has been solved exactly 
in one and two dimensions. However, 
for higher dimensions,
there exist extensive perturbative analyses
based on renormalization group methods and
numerical investigations with Monte Carlo methods.

It was traditionally believed that universal scaling behaviour
exists only in or near thermodynamic equilibrium.
Recently, it has been argued 
theoretically \cite{jan89}
that some dynamic systems already exhibit universal scaling behaviour
in the macroscopic short-time 
region of their dynamic evolution.
The main point is that universality and scaling 
emerge after a microscopic time scale $t_{\rm mic}$,
which is sufficiently large in the microscopic sense
but still short in the macroscopic sense.
This statement has been proven 
valid in a series of numerical simulations
of various statistical systems.
More interestingly,
as well as
the new critical exponent $\theta$, describing
the critical initial increase of the magnetization,
and the dynamic critical exponent $z$,
short-time critical dynamics provides a measure of
the static critical exponents and of
the critical temperature.
This dynamic approach is free of critical slowing down
since the spatial correlation length is still small
within the short-time regime, even at or near 
the critical point\footnote{
For a recent review see Ref.~\protect\cite{zhe98}.}.

Up to now, systematic numerical simulations
of the short-time critical dynamics 
have been carried out mainly in two-dimensional
systems \cite{zhe98}. 
For the three-dimensional Ising model,
the power-law decay of the magnetization
starting from an ordered state
has been simulated \cite{sta96a,sta97},
the new exponent $\theta$ and the dynamic exponent 
$z$ have also been obtained \cite{li94,gra95,gro95}. 
However, a complete understanding of
the short-time dynamic behaviour of 
the three-dimensional Ising model is still
necessary, since it 
is a very important model. Especially, a systematic test of
the short-time dynamic approach to the determination
of all the critical exponents and the critical temperature
in a three-dimensional system is important.
 
In this paper we report a comprehensive
investigation of the short-time critical dynamics of
the three-dimensional Ising model.
A power-law behaviour of the autocorrelation,
the second moment and the Binder cumulant is observed,
in addition to that of the magnetization.
The results support fully the short-time dynamic scaling.
For the first time we extract the critical temperature
and the static exponents $\nu$ and $\beta$ 
from the short-time scaling behaviour.
Our 
results for the static exponents and critical temperature
agree well with those obtained in the extensive studies 
at 
thermodynamic equilibrium.

In the next section, a scaling analysis
of the short-time critical dynamics is given.
Numerical results are presented in Sec.~III.
The last section contains the summary and discussion.

\section{Scaling relations}

Using renormalization group methods,
Janssen, Schaub and Schmittmann \cite{jan89} have 
shown
that far from equilibrium,
in a macroscopic short-time regime 
of the dynamical evolution, 
there already emerges universal scaling behaviour
in the $O(N)$ vector model.
The relaxation process considered is one of a system
initially in a high-temperature state, 
suddenly quenched to the critical temperature $T_c$
and evolving with dynamics of model A.
For an initial state with a non-zero but small
magnetization $(m_0\ll1)$, a generalized dynamic 
scaling form has been derived 
with an $\epsilon$-expansion for the $O(N)$ vector model,
\be
M^{(k)}\;(t,\tau,L,m_0)=b^{-k\beta/\nu}\;
   M^{(k)}(b^{-z}t,b^{1/\nu}\tau,b^{-1}L,b^{x_0}m_0).
\label{sc}
\ee
In Eq.~(\ref{sc}), $M^{(k)}$ is the $k$th moment of the 
magnetization,
\be
M^{(k)}(t) = \frac{1}{L^3}\;
\left\langle \left(\sum_i S_i(t)\right)^{k} 
\right\rangle,
\label{mk}
\ee
$t$ is the time of the dynamical relaxation, $L$ is the 
lattice size,
\be
\tau = \frac{T-T_c}{T_c}
\ee
is the reduced temperature,  and
$b$ is a spatial rescaling factor. The quantity
$x_0$ is a new independent critical exponent, 
the scaling dimension of the initial magnetization 
$m_0$. The interesting and important point
is that  
the critical exponents 
$\beta$, $\nu$ and $z$ in Eq.~(\ref{sc}) are exactly
those usually defined in equilibrium. 
These exponents can thus be extracted
from the short-time critical dynamics.

Equation~(\ref{sc}) provides
the time evolution of the magnetization, with
$k=1$. Taking $b=t^{1/z}$ for the spatial scaling factor 
gives
\bea
M(t,\tau,m_0) &=& t^{-\beta/\nu z}\;M(1,t^{1/\nu z}\tau,t^{x_0/z}m_0)
\label{m0ex}\\
&\sim&m_0\,t^{(x_0-\beta/\nu)/ z}\;F(t^{1/\nu 
z}\tau) +    {\cal O}([t^{x_0/z}m_0]^2)\;. 
\nonumber
\eea
The expansion has been performed
with respect to the small quantity 
$t^{x_0/z}m_0$ and 
$M(t,m_0=0)=0$ has been used. It has been implicitly
assumed
that $L$ is sufficiently large.
Exactly at the critical point 
$(\tau=0)$, Eq.~(\ref {m0ex}) predicts a
power-law behaviour of the magnetization in the 
short-time region,
\be
M(t) \sim t^\theta, \quad \theta=\left(x_0-\frac{\beta}{\nu}\right)\;\frac{1}{z}\ .
\label{th}
\ee
Up to now, analytical calculations for the $O(N)$ vector
model and numerical simulations for a variety of
statistical systems show that
 $\theta>0$, i.e. the magnetization undergoes 
 an initial {\it increase}.
 This is a very prominent phenomenon in the short-time critical dynamics.

Now we consider the case $m_0=0$.
Using Eq.~(\ref{sc})
for the second moment of the 
magnetization
at the critical temperature gives
\be
M^{(2)}(t) \sim t^{-2\beta/\nu z}\; 
M^{(2)}(1,t^{-1/z}L).
\label{m2}
\ee
In the beginning of the time evolution the spatial 
correlation length
is still small, even at the critical point. 
Thus it can be deduced
that $M^{(2)}(t,L)\sim L^{-d}$, where $d$ is the 
dimension of the
system. 
Taking this into account,
Eq.~(\ref {m2})
yields a power-law behaviour
\be
M^{(2)}(t) \sim t^{c_2}, \qquad 
c_2=\left(d-2\frac{\beta}{\nu}\right)\;\frac{1}{z}.
\label{c20}
\ee
An analysis of the
autocorrelation (for $m_0=0$) 
\be
A(t) = \frac{1}{N}\;
\left\langle \sum_i S_i(t)S_i(0) \right\rangle
\label{a}
\ee
shows that it obeys a power-law \cite{jan92}
\be
A(t)\sim t^{-c_a},\qquad c_a=\frac{d}{z}-\theta.
\label{ca}
\ee

In summary,
 simulations of the dynamic system starting from small 
or zero initial magnetization, at or near the critical 
point, allow the quantities
$\theta$, $c_2$ and $c_a$ to be measured 
and thus 
a determination of the critical exponents 
$\theta$, $\beta$/$\nu$ and $z$ separately. 
In principle, 
the critical temperature itself can also be determined 
from the location of
the optimal power-law behaviour of the
magnetization within the critical region
\cite {zhe98,sch96}. 
However,
a similar but more accurate determination
of the critical temperature will be presented below.
This is also the case for the determination of the exponent
$\nu$ from the derivative of $M$ with respect to  $\tau$.

\vspace{0.5cm}

In the above considerations the dynamic relaxation 
process was assumed to start from 
a disordered state with vanishing or small 
magnetization $m_0$. 
Another interesting and important process
is the dynamic relaxation from a completely ordered state.
The initial magnetization being exactly 
at its fixed point $m_0=1$,
a scaling form
\begin{equation}
M^{(k)}\;(t,\tau,L)=b^{-k\beta/\nu}\;
   M^{(k)}(b^{-z}t,b^{1/\nu}\tau,b^{-1}L)
\label{sc1}
\end{equation}
is expected.
This scaling form looks the same as the dynamic scaling form
in the long-time regime, however, it is now assumed 
already valid in the macroscopic short-time regime.

For the magnetization itself, 
$b=t^{1/z}$ yields
\begin{equation}
M(t,\tau) = t^{-\beta/\nu z}\;M(1,t^{1/\nu z}\tau).
\label{m0ex1}
\end{equation}
This leads to the power-law behaviour
\be
M(t)\; \sim \; t^{-c_1}, \qquad c_1=\frac{\beta}{\nu z}
\label{c1}
\ee
at the critical point ($\tau=0$).
For small but nonzero $\tau$, the power-law behaviour of the magnetization
will be modified by the scaling function 
$M(1,t^{1/\nu z}\tau)$, thus allowing for 
a determination of
the critical temperature \cite {zhe98}.
Taking the derivative with respect to $\tau$
on both sides of Eq.~(\ref{m0ex1}) and
fixing 
$b=t^{1/z}$ again, gives the logarithmic derivative of the
magnetization 
\be
\partial_\tau \ln\; M(t,\tau)|_{\tau=0}\; \sim\; t^{-c_{\ell 1}},
\qquad  c_{\ell 1}=\frac{1}{\nu z}.
\label{cl1}
\ee

Here, unlike the relaxation from
a disordered state, 
the average magnetization is not zero.
A Binder cumulant $U(t)$ can be obtained using the magnetization
and its second moment. Finite size scaling
shows that
\be
U(t)=\frac{M^{(2)}}{(M)^2}-1 \; \sim \;  t^{c_U},\qquad 
   c_U=\frac{d}{z}\ .
\label{cu}
\ee

Thus, 
the short-time behaviour of the dynamic relaxation
starting from a completely ordered state 
is also sufficient to
determine all the critical exponents
$\beta$, $\nu$ and $z$ as well as the critical temperature.
In practical simulations, these measurements of
the critical exponents and critical temperature 
are usually better in quality than those from a relaxation
process starting from a disordered state.

\section{Numerical results}

We have performed simulations on three-dimensional lattices of
linear sizes $L=32$, $64$ and $128$ (in a particular case $L=256$),
starting 
either from an ordered state
or from a high-temperature state  
with zero or small initial magnetization. 
In the latter cases the inital magnetization
 has been prepared by 
flipping in an ordered state a definite number of spins 
at randomly chosen sites in order to 
get the desired small value of $m_0$. Starting from
the initial state, the system has been 
updated by a heat bath Monte Carlo algorithm. A unit  in time
is defined as a complete update of all the spins in the lattice. 
Simulations have been performed, depending on the inital magnetization,
up to $t=1000$. The preparation of the inital magnetization and the update
up to the maximal time has been repeated $1000$ or $5000$
times, depending on the lattice size (much less for $L=256$),
and the magnetization, the
second moment of the magnetization (\ref{mk})
or the autocorrelation (\ref{a}) has been measured.
In general, several runs of this kind have been
performed to estimate the statistical error.
The simulations used the value $K_c=0.22166$ 
for the critical point, taken from Ref.\cite{fer91}.
Simulations have also been carried out
in the neighbourhood of $K_c$ 
to extract the critical point and the critical exponent $\nu$.

\subsection{Evolution from a disordered state}

\noindent{\bf Magnetization}\newline
At the beginning of the time evolution,
for sufficiently small $t^{x_0/z}m_0$, 
the magnetization undergoes
a power-law initial increase.
For $m_0$ and $t$ not too small,
the power-law behaviour will be modified.
This may be called `finite $m_0$ effect'.
The possibility of a
finite size effect also has to be evaluated.
Furthermore, scaling form (\ref{sc}) is
strictly valid only in the limit $m_0=0$.
However, practical measurements can only be carried out
for finite $m_0$ and the measured exponent $\theta$
may show a weak dependence on $m_0$.
The data must thus be extrapolated for $m_0=0$.

\begin{figure}\centering
\epsfclipoff
\fboxsep=0pt
\setlength{\unitlength}{0.75cm}
\begin{picture}(12,8)(0,0)
\put(0,0){{\epsfxsize=9cm \epsfbox{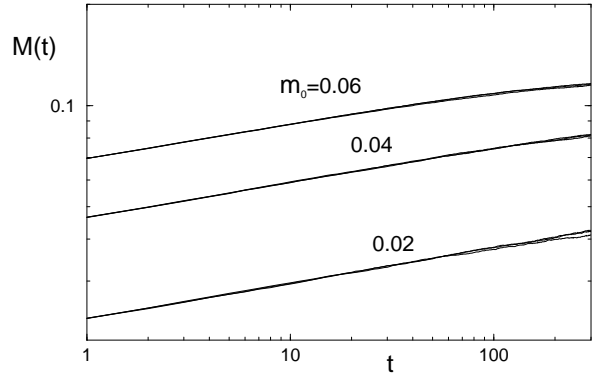}}}
\end{picture}
\caption{ 
Time evolution of the magnetization $M(t)$ 
for three values of the 
initial magnetization $m_0$ for $L=128$.
}
\label{f0a}
\end{figure}

We have measured the magnetization $M(t)$
for $m_0=0.02$, $0.04$ and $0.06$
for the largest lattice, $L=128$.
For $m_0=0.02$ we have 
performed four runs, in the other cases three runs up to $t=300$.
Fig.~\ref{f0a} shows on a log-log scale
a pronounced power-law behaviour starting at the
onset of the time evolution. A
finite $m_0$ effect is observed at the larger times.

In order to determine more clearly the region where the power-law
is strictly fulfilled, 
we divided the time scale
into non-overlapping intervals $\Delta_q(t)=[t,qt]$,
with $q<2$,
and measured the exponent $\theta(t)$ separately in 
each interval.
Since the intervals contain more points
at later times where the fluctuations are larger,
the errors in each 
interval are comparable. 
Already from the second bin the slope shows a stable
behaviour. 
For $m_0=0.02$ slope 
is quite stable up to
$t=100$.
Therefore  $\theta$ was calculated
from a least-squares fit in the time interval 
$[2,100]$.
\begin{table}\centering
\begin{tabular}{|c|c|c|c|c|}
$m_0$       & \ 0  & \ 0.02 & \ 0.04 &\ 0.06\\
\hline
 $\theta$      & \ 0.108(2)  & \ 0.1059(20)& \ 0.1035(4) &\ 0.1014(5)\\
\end{tabular}
\caption{
The exponent $\theta$ measured for $L=128$ for
different values of the initial magnetization $m_0$.
The value $\theta(m_0=0)$ is the result of an extrapolation 
(see text).
}
\label{t1}
\end{table}
\noindent 
Similarly, 
we chose the interval $[2,50]$ 
for $m_0=0.04$. For $m_0=0.06$ the restrictive
interval $[2,15]$ was used.
Table~\ref{t1} presents the results.
From these results, an
extrapolation to $m_0=0$ yields
$\theta=0.108(2)$. 
Careful analysis of the data for 
$L=64$ shows that the finite size effect in $L=128$
is already negligible.

\vspace{0.5cm}

\begin{figure}\centering
\epsfclipoff
\fboxsep=0pt
\setlength{\unitlength}{0.75cm}
\begin{picture}(12,12)(0,0)
\put(0.,5.2){{\epsfxsize=8cm \epsfbox{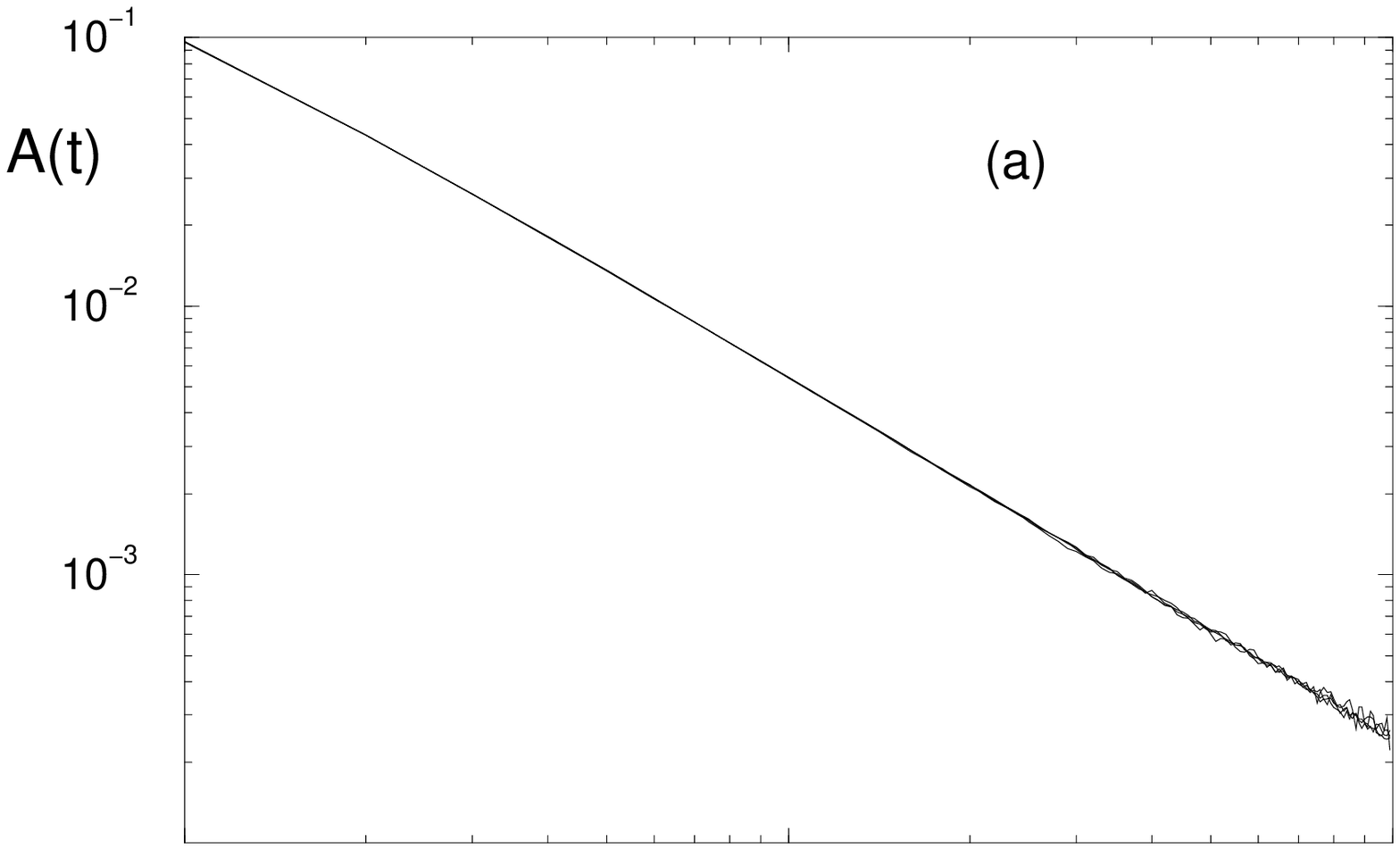}}}
\put(0.,-0.2){{\epsfxsize=8cm \epsfbox{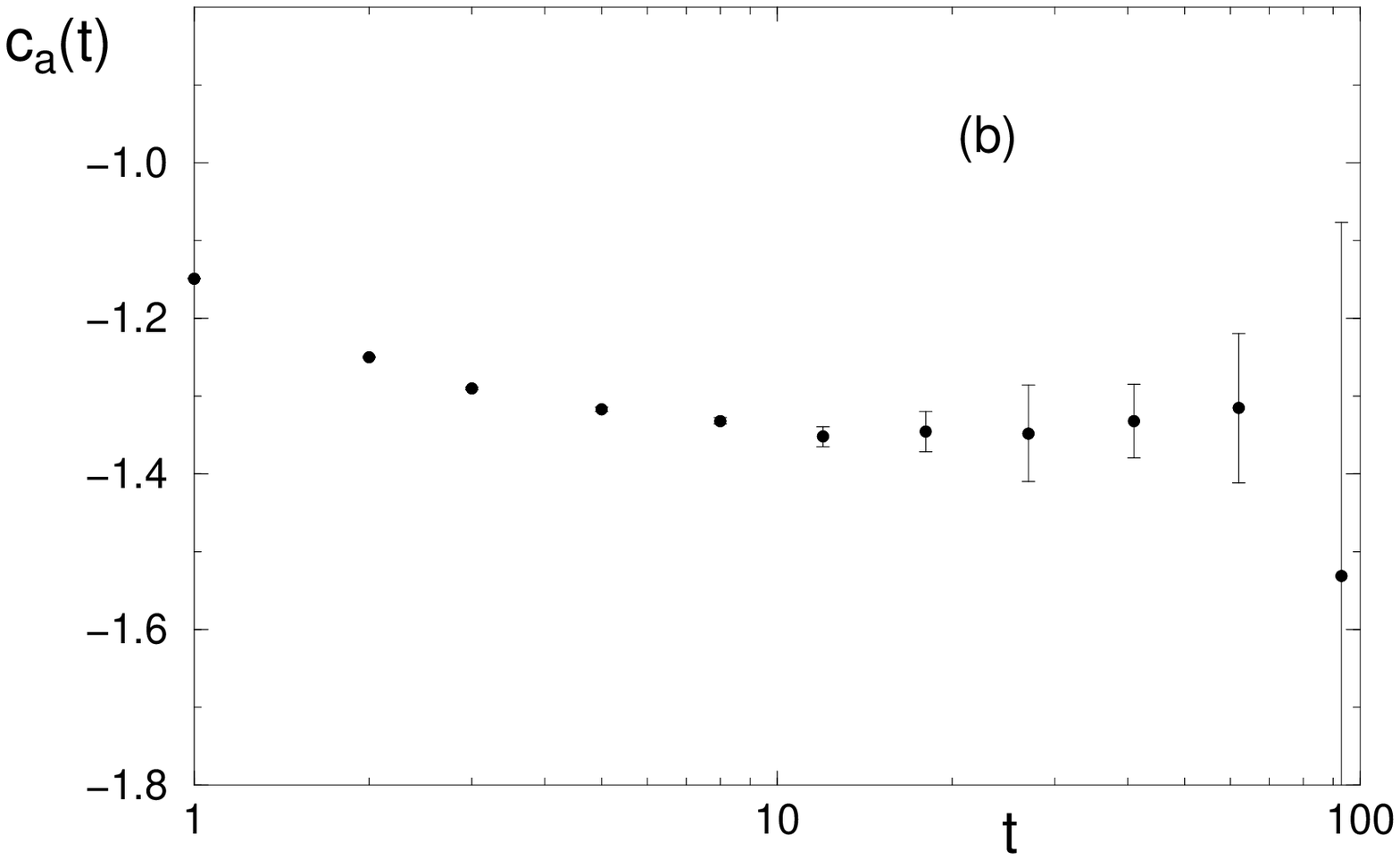}}}
\end{picture}
\caption{
(a) Autocorrelation $A(t)$ and 
 (b) the slope $c_a(t)$ of the autocorrelation 
function for $L=128$, in the intervals 
$\Delta_{1.5}(t)=[t,1.5t]$. 
}
\label{f0c}
\end{figure}

\noindent{\bf Autocorrelation}\newline
Measurements of the autocorrelation function suffer from large 
fluctuations in the region of large $t$, 
since it decays by nearly three orders of magnitude
between $t=0$ and $t=100$. Fig.~\ref{f0c}(a) shows 
results for four runs for $L=128$
on a log-log scale. 
For the lattice size $L=64$ three runs were performed.

As described above, we have divided the time scale into
non-overlapping intervals $\Delta_{1.5}(t)$ and 
calculated $c_a(t)$ in each interval.
This is exemplified for $L=128$ 
in Fig.~\ref{f0c}(b). 
On this plot 
$c_a(t)$ is rather constant after the first few bins, while
large fluctuations start at $t>70$. 
Thus, 
the interval $t=[10,70]$ is used for $L=128$ and $64$
in order to obtain $c_a$.
The results are given in table~\ref{t2}.
With no apparent dependence on lattice size, 
the two results were simply
averaged to estimate $c_a$ for $L=\infty$
(second column).

\vspace{0.5cm}

\begin{figure}\centering
\epsfclipoff
\fboxsep=0pt
\setlength{\unitlength}{0.75cm}
\begin{picture}(12,8)(0,0)
\put(0.5,.2){{\epsfxsize=8cm \epsfbox{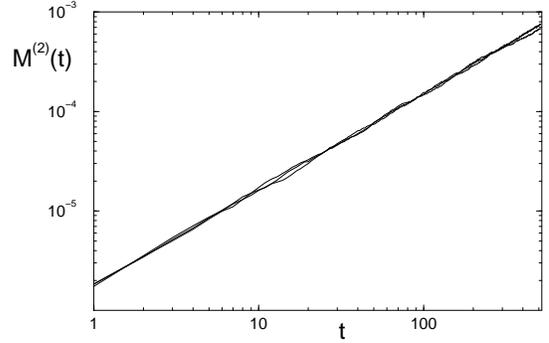}}}
\end{picture}
\caption{
Time evolution of the second moment of the
magnetization, $M^{(2)}(t)$, for $L=128$.
}
\label{f0d}
\end{figure}

\noindent{\bf Second moment of the magnetization}\newline
The second moment of the magnetization has been simulated
for $L=128$ in three runs up to $t=512$, 
and for $L=64$  in three runs each
up to $t=100$. Fig.~\ref{f0d} shows $M^{(2)}(t)$ 
for $L=128$ on a log-log scale.
This curve as well as the slopes
in bins show slightly larger
fluctuations at higher $t$, but no departure from
a power-law. Therefore, the 
full intervals $t=[2,512]$
for $L=128$ and $t=[2,100]$ for $L=64$
were used to compute the values of $c_2$ 
reported in table~\ref{t2}. 
For $L=\infty$, the estimate
$c_2=0.970(11)$ provided is obtained by averaging the results
for $L=64$ and $L=128$,
since no dependence on the lattice size
can be seen.

\subsection{Evolution from an ordered state}

For the determination of the critical exponents and 
critical temperature, 
the dynamic relaxation process starting from
an ordered state  has been proven 
advantageous over that 
from a high-temperature initial state. 
Indeed, the magnetization 
only decreases slowly in time from $m_0=1$.
Furthermore, the magnitude
of the magnetization in the short-time regime is 
large, therefore  
statistical fluctuations are less prominent. 
We have measured 
the magnetization $M(t)$ and the second moment 
$M^{(2)}(t)$ for lattice sizes $32$, $64$ and $128$,
and, with less statistics, for $L=256$. 
Measurements have been performed up to
$t=512$ for $L=128$, and up to $t=1000$ for
the magnetization only. 
For the large
lattice size $L=256$ an average over only 20 samples
has been performed whereas
for $L=128$ and $64$, over 1000 samples were taken for
each run, and two to four runs performed in
order to estimate the statistical error. 
For $L=32$, $5000$ samples were collected.

\vspace{0.5cm}

\begin{figure}\centering
\epsfclipoff
\fboxsep=0pt
\setlength{\unitlength}{0.75cm}
\begin{picture}(12,7)(0,4)
\put(0,4.){{\epsfxsize=8cm \epsfbox{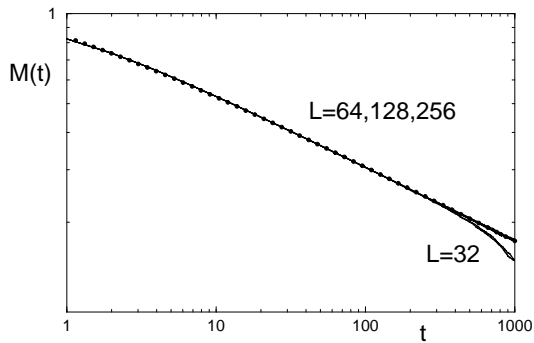}}}
\end{picture}
\caption{
Time evolution of the
magnetization $M(t)$ for $m_0=1$ 
for $L=256$, $128$, $64$, and $32$. 
The points are obtained from
the best fit to
(\protect\ref{pf}).
}
\label{f1a}
\end{figure}

\noindent{\bf Magnetization}\newline
Fig.~\ref{f1a} shows the evolution of the magnetization 
for all four lattice sizes 
on a log-log scale.
It is interesting that for $L=256$ down to $L=64$ 
the measurements completely overlap up to $t=1000$. 
From Eq.~(\ref {c1}),
the slope of the curves provides a measure
of $\beta/\nu z$. However, careful analysis
reveals that 
the slope decreases weakly with increasing time.
This suggests that a correction to scaling should be
considered in order to obtain accurate results.
In comparison to the two-dimensional
Ising model, the correction to scaling 
seems somewhat larger in three dimensions \cite{zhe98}. 
We have found that
the best correction to scaling is given by the
{\it Ansatz} 
\be
M(t)=a\;t^{-c_1}\;e^{-\gamma t^{-\delta}}\;.
\label{pf}
\ee
Least-squares fits were performed
in the interval $t=[1,1000]$
for $L=64$, $128$ and $256$.
The resulting fit parameters,
averaged over the three lattice sizes
and over the two runs for each lattice size,
are $\ln (a)=0.0033(10)$, 
$c_1=0.2533(7)$, $\gamma=0.1874(19)$, and $\delta=0.479(39)$.
Table~\ref{t2} reports the values for $c_1$.
Calculated values for $M(t)$ using (\ref{pf})
are also provided on Fig.~\ref{f1a} (points) and show good agreement.
Without the correction to scaling,
the measured exponent $c_1$ would be one to two percent
smaller.

\begin{figure}\centering
\epsfclipoff
\fboxsep=0pt
\setlength{\unitlength}{0.75cm}
\begin{picture}(12,8)(-0.7,0)
\put(0,0){{\epsfxsize=8cm \epsfbox{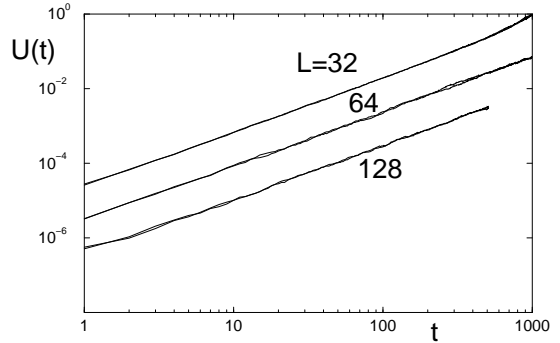}}}
\end{picture}
\caption{
Time evolution of the Binder cumulant $U(t)$ for $m_0=1$
and lattice sizes $L=128$, $64$, and $32$.
}
\label{f1b}
\end{figure}

\noindent{\bf Binder cumulant}\newline
The second moment of the magnetization has been measured up to
$t=512$ for $L=128$, and up to $t=1000$ for $L=64$ and $32$.
A plot of the Binder cumulant $U(t)$, defined in Eq.~(\ref{cu}), 
is shown on log-log scale
in Fig.~\ref{f1b}. 
The curves show that
for $L=32$ the power-law behaviour prevails only up to $t\sim100$,
while for the larger lattices it remains
up to the maximal time. 
An analysis of the slope measured in the intervals
$\Delta_{1.5}(t)$ shows that the exponent $c_U(t)$  can be obtained in the
interval $t=30$ up to the maximal time for $L=64$ and $128$.
For $L=32$ there is a plateau only over $t=[10,100]$.
Hence the results for this lattice size are not reported.
The slope $c_U$ is calculated 
in the intervals $t=[30,1000]$ for
$L=64$ and $t=[30,512]$ for $L=128$ and given in
table~\ref{t2}. 
Within errors they are consistent and their average
is reported for $L=\infty$ in table~\ref{t2}.
The error estimates are only based on statistics
over a limited number of runs and exclude possible systematic 
contributions.

\begin{figure}\centering
\epsfclipoff
\fboxsep=0pt
\setlength{\unitlength}{0.75cm}
\begin{picture}(12,6)(0,0)
\put(.5,-0.2){\epsfxsize=8cm \epsfbox{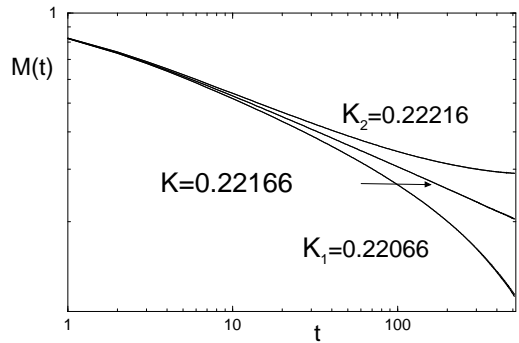}}
\end{picture}
\caption{
Time evolution of the 
magnetization  $M(t)$ for $m_0=1$
for three values of the inverse temperature
$K_1=0.22066$, $K=0.22166$, and $K_2=0.22266$.
}
\label{f1c}
\end{figure}
\begin{figure}\centering
\epsfclipoff
\fboxsep=0pt
\setlength{\unitlength}{0.75cm}
\begin{picture}(12,6)(0,0)
\put(0,-0.4){\epsfxsize=8cm \epsfbox{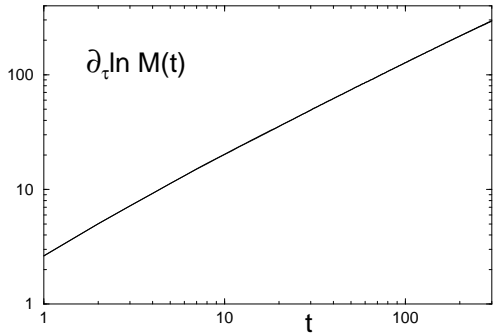}}
\end{picture}
\caption{
Logarithmic derivative of the magnetization with respect to
$\tau$,
obtained from a quadratic interpolation between the three curves shown in
Fig.~\protect\ref{f1c} taken at $K=0.22166$.
}
\label{f7}
\end{figure}

\vspace{-.3cm}
\noindent{\bf Critical temperature and $c_{\ell1}$}\newline
Figure~\ref{f1c} shows the
magnetization  $M(t)$ for $m_0=1$
for three values for the inverse temperature
$K_1=0.22066$, $K=0.22166$ and $K_2=0.22266$, on a
log-log scale.
The critical point $K_c$ can be estimated by searching for 
the best power-law behaviour
with $K$ between $K_1$ and $K_2$.
Namely, the best straight-line fit to curves obtained
by quadratic interpolation for
$K_1<K<K_2$ is sought.
Restricting the interval to $t=[30,512]$, the result $K_c=0.22170(4)$
is obtained.
The quadratic interpolation also provides 
the logarithmic derivative of the magnetization with respect to $\tau$ in
Eq.~(\ref{cl1}). 
This is shown in Fig.~\ref {f7}.
The slope over the interval $[30,300]$ provides 
$c_{\ell1}=0.774(1)$. The errors are based on the statistics 
of comparing only
two data sets.
A correction to scaling has not been considered.
This could result in slightly larger errors.

\begin{table}\centering
\begin{tabular}{|l|l|l|l|}
       & \ $L=\infty$  & \ $L=128$ & \ $L=64$ \\
\hline
 $c_a$      &\  1.362(19)
            &\  1.358(19)
            &\  1.366(20)  \\
\hline
 $c_2$      & \ 0.970(11)  & \ 0.966(16) & \ 0.975(3) \\
\hline
\hline
 $c_1$      & \ 0.2533(7) & \ 0.2539(1) & \ 0.2527(8) \\
\hline
 $c_U$      & \ 1.462(12) & \ 1.453(17) & \ 1.471(4) \\
\hline
 $K_c$        & \ 0.22170(4) & \ 0.22170(4) & \ -- \\
\hline
 $c_{\ell1}$        & \ 0.774(1) & \ 0.774(1) & \ --\\
\end{tabular}
\caption{
Exponents $c_a$ of the autocorrelation function and $c_2$ of
the second moment of the magnetization for different lattice
sizes obtained from a disordered initial state 
for $m_0=0$ (upper part). Exponents
$c_1$ of the power-law decrease of the magnetization and $c_U$ for the
cumulant, as well as inverse temperature $K_c$
and exponent
$c_{\ell1}$ for the logarithmic derivative of the magnetization
measured from an ordered initial state (lower part).
For $c_1$, the additional result $c_1=0.2534(4)$
obtained from two runs with $L=256$ has been included (see text). 
}
\label{t2}
\end{table}

\subsection{Critical exponents\label{crex}}
We now proceed to determine the critical exponents.
Using the 
relaxation from
a high-temperature state,
we have measured the exponent $\theta$ directly from the
critical initial increase in magnetization (\ref{th}), 
and the exponents 
$c_2$ and $c_a$ from the power-law behaviour of 
the second moment (\ref{c20})
and the autocorrelation (\ref{ca}). 
Similarly, from measurements of the power-law
behaviour with an {\it ordered} initial state we have obtained $c_1$,
$c_{\ell1}$, and $c_U$ from the magnetization (\ref{c1}), 
its logarithmic derivative (\ref{cl1}) and the Binder cumulant (\ref{cu}).
At this point, it is useful to recall 
the scaling relations:
$$
\begin{array}{rl rl}
\displaystyle c_2=&\frac{d}{z}-2\frac{\beta}{\nu z},& \hspace*{1cm} c_1=&\frac{\beta}{\nu z},
\\ \\
\displaystyle c_a=&\frac{d}{z}-\theta, & c_U=&\frac{d}{z},\\ \\
\displaystyle        &       &c_{\ell1}=&\frac{1}{\nu z}.
\end{array}
$$

Our measurements of the exponent $c_1$ 
are very accurate, while actual errors on $K_c$ and
$c_{\ell1}$ could be somewhat larger than those
 given in table~\ref {t2}.
The value obtained for $c_1$ agrees well with previous measurements
\cite {sta96a,sta97}, 
although the approaches to correct for scaling differ.
Our procedure is essentially similar to that used 
in Ref.~\cite {ito93}.
The dynamic exponent
$z$ can be estimated independently from $c_U$.
Since the Binder cumulant (\ref{cu}) is constructed from
the magnetization and its second moment,
the estimates are usually better
than for a Binder cumulant constructed from the second
and the fourth moments, typically used
in the relaxation from
a disordered state or in equilibrium.
The exponent $z$ can also be extracted
from $c_a+\theta$ and 
$c_2+2c_1$. 
Table~\ref{t3} lists the results.
These concur within statistical errors.
Table~\ref{t4} reports the average of the three values.
With $z$, and from $c_1$ and $c_{\ell1}$, 
the exponents $\beta/\nu$ and $\nu$ can be estimated.
Results for all critical exponents and the critical 
point $K_c$ are given in table~\ref{t4}.

\begin{table}\centering
\begin{tabular}{|c|l|l|}
       & \ $d/z$  & \ $z$  \\
\hline
 $c_a+\theta$   & 1.470(13) &  \    2.041(18) \\
\hline
 $c_U$          & 1.462(12) &  \    2.052(17) \\
\hline
 $c_2+2c_1  $   & 1.4766(8) &  \    2.032(11) \\
\end{tabular}
\caption{
Determination of $z$ from three independent measurements of $d/z$.
}
\label{t3}
\end{table}


Our final result for the exponent $\theta$ given in table~\ref {t1},
$\theta=0.108(2)$ is consistent with an early estimate using a small 
lattice \cite{li94} but slightly larger than $\theta=0.104(3)$ 
obtained from damage spreading 
\cite {gra95}. It is not clear whether this difference 
comes from statistical or systematic errors.
From equilibrium dynamics, the dynamic exponent $z=2.04(3)$ 
has been extracted
\cite{wan91}.
Values of $z=2.032(4)$ \cite{gra95}
and $z=2.04(1)$ \cite{gro95} have been obtained recently from 
damage spreading and of $z=2.05(2)$ \cite{sta96a}
and $z=2.04(2)$ \cite{sta97} from large scale
simulations of critical relaxation, 
starting from an ordered state. Our value
$z=2.042(6)$
is consistent with all of them.
 For the ratio $\beta/\nu$ and $\nu$,
recent analytic calculations on 
the base of a renormalization--group expansion (in equilibrium) yield
$\beta=0.327$, $\beta/\nu=0.518$ and $\nu=0.631$ \cite{gui80,gui87,ant98},
while numerical investigations in equilibrium by
Ferrenberg and Landau \cite{fer91}  
yield $\beta=0.3258(44)$,
$\beta/\nu=0.518(7)$ and $\nu=0.6289(8)$. 
Ref.~\cite{fer91} provides a good review of earlier numerical values.
Investigations using the cluster algorithm have yielded
$\nu=0.6301(8)$ and $\beta=0.3267(10)$ \cite{blo95}
and $\beta=0.3269(6)$ \cite{tal96}.
In \cite{ito93}, the value 
$\nu=0.6250(25)$ has been extracted from measuring the interface energy.
Our value of $\beta/\nu=0.517(2)$ is accurate
but the value of $\nu$ is slightly larger.
Our estimate of $K_c$ is consistent with $K_c=0.2216595(26)$
in Ref.~\cite{fer91}, $K_c=0.2216546(10)$ \cite{blo95},
or $K_c=0.2216544(3)$ \cite{tal96}.
Similar results have been quoted in Refs.~\cite{bai92,ito91,ito90}.

\begin{table}
\begin{tabular}{r|l}
$\theta= 0.108(2)$ \hspace{.88cm} &  0.104(3)\cite{gra95} \\
\hline
$z=2.042(6)$ \hspace{.88cm} & 
   \begin{tabular}{l}
   2.04(3)\cite{wan91},\ 2.032(4)\cite{gra95},\\
   2.04(1)\cite{gro95},\ 2.05(2)\cite{sta96a},\\
   2.04(2)\cite{sta97} 
   \end{tabular} \\
\hline
$\beta/\nu=0.517(2)$ \hspace{.88cm} &  0.518(7)\cite{fer91},\
0.5185(16)\cite{blo95}\\
\hline
$\nu=0.6327(20)$ \hspace{.88cm} &  
    \begin{tabular}{l}
    0.6289(8)\cite{fer91},\ 0.6250(25)\cite{ito93} \\
    0.6301(8)\cite{blo95}
    \end{tabular}\\
\hline
$\beta=0.3273(17)$ \hspace{.88cm} &  
    \begin{tabular}{l}
    0.3258(44)\cite{fer91},\ 0.3267(10)\cite{blo95}\\
    0.3269(6)\cite{tal96}
    \end{tabular}\\
\hline
$K_c=0.22170(4)$ \hspace{.88cm} &  
    \begin{tabular}{l}
    0.2216595(26)\cite{fer91}\\
    0.2216546(10) \cite{blo95}\\
    0.2216544(3) \cite{tal96}
    \end{tabular}
\end{tabular}
\caption{
Final results for all critical exponents and the critical point 
$K_c$ (left). 
Results from earlier investigations 
discussed in the text have been collected
in the right column.}
\label{t4}
\end{table}

\section{Summary and Discussion}

In the previous sections we have reported comprehensive
Monte Carlo simulations of the short-time critical dynamics 
for the three-dimensional Ising model. 
Starting from a
high-temperature initial state,
 the magnetization, its second moment
and the autocorrelation have been measured at the critical point.
Similarly, we have studied the 
behaviour of the magnetization, its derivative
with respect to the temperature and the Binder cumulant
 for a completely ordered initial state.
Theoretically one expects a power-law for all these observables
within the short-time regime at the critical point.
This is indeed observed in our numerical simulations.
All the dynamic exponents and static exponents as well as
the critical temperature have been determined.
The results are consistent and strongly support
a full dynamic scaling in the short-time regime
of the dynamic evolution.
The values of the dynamic exponent $z$ and the static
exponents $\beta/\nu$ and $\nu$ are independent of
initial conditions and agree well with 
 those measured in equilibrium. 
Unlike non-local cluster algorithms, the short-time
dynamic approach
studies the dynamic properties of the original local dynamics.
Our measurements of the static exponents are comparable with large
scale simulations in equilibrium.
All this indicates that the dynamic measurements of the
critical exponents are promising. 
The present work could possibly be extended
to the diluted or the random field Ising
models.

\vspace{0.4cm}

\noindent{\bf Note added}\newline
Recently, the results $\nu=0.6298(5)$ \cite{has98}
and $\beta/\nu=0.518(1)$ \cite{cho98} have been reported.

\newpage


\begin{thebibliography}{10}

\bibitem{jan89}
{H. K. Janssen, B. Schaub and B. Schmittmann}, Z. Phys. {\bf {B 73}} (1989)
  539.

\bibitem{zhe98}
B. Zheng, Int. J. Mod. Phys. {\bf B 12} (1998) 1419.

\bibitem{sta96a}
{D. Stauffer and R. Knecht}, Int. J. Mod. Phys. {\bf {C7}} (1996) 893.

\bibitem{sta97}
D. Stauffer, Physica {\bf A 244} (1997) 344.

\bibitem{li94}
{Z.B. Li, U. Ritschel and B. Zheng}, J. Phys. A: Math. Gen. {\bf {27}} (1994)
  L837.

\bibitem{gra95}
P. Grassberger, Physica {\bf {A 214}} (1995) 547.

\bibitem{gro95}
U. Gropengiesser, Physica {\bf A 215} (1995) 308.

\bibitem{jan92}
{H. K. Janssen},  in {\em {From Phase Transition to Chaos}}, edited by {G.
  Gy\"orgyi, I. Kondor, L. Sasv\'ari and T. T\'el, Topics in Modern Statistical
  Physics} (World Scientific, Singapore, 1992).

\bibitem{sch96}
L. Sch{\"u}lke and B. Zheng, Phys. Lett {\bf {A 215}} (1996) 81.

\bibitem{fer91}
A.~M. Ferrenberg and D.~P. Landau, Phys. Rev. {\bf {B 44}} (1991) 5081.

\bibitem{ito93}
N. Ito, Physica {\bf {A 196}} (1993) 591.

\bibitem{wan91}
S. Wansleben and D.~P. Landau, Phys. Rev. {\bf {B 43}} (1991) 6006.

\bibitem{gui80}
J.-C.~L. Guillou and J. Zinn-Justin, Phys. Rev. {\bf B 21} (1980) 3976.

\bibitem{gui87}
J.-C.~L. Guillou and J. Zinn-Justin, J. Phys. (Paris) {\bf 48} (1987) 19.

\bibitem{ant98}
S.~A. Antonenko and A.~I. Sokolov, {\em Critical exponents for $3D$
  $O(N)$--symmetric model with $n>3$}, Electrotechnical University, St.
  Petersburg, 1998, preprint hep-th/9803264.

\bibitem{blo95}
{H.W.J. Bl{\"o}te, E. Luijten and J.R. Heringa}, J. Phys. A: Math. Gen. {\bf
  28} (1995) 6289.

\bibitem{tal96}
{A.L. Talapov and H.W.J. Bl{\"o}te}, J. Phys. A: Math. Gen. {\bf 29} (1996)
  5727, cond-mat/9603013.

\bibitem{bai92}
{C.F. Baille, R. Gupta, K.A. Hawick and G.S. Pawley}, Phys. Rev. {\bf B 45}
  (1992) 10438.

\bibitem{ito91}
N. Ito and M. Suzuki, Journ. Phys. Soc. Jpn. {\bf 60} (1991) 1978.

\bibitem{ito90}
N. Ito,  in {\em AIP Conf. Proc.}, edited by C.-K. Hu (AIP, New York, 1990),
  Vol.~248, p.\ 136, computer Aided Statistical Physics (Taipei, Taiwan,1991).

\bibitem{has98}
{M. Hasenbusch, K. Pinn and S. Vinti}, {\em Critical exponents of the 3D Ising
  universality class from finite size scaling with standard and improved
  actions}, 1998, hep-lat/9806012.

\bibitem{cho98}
{Y.S. Choi, J. Machta, P. Tamayo, and L.X. Chayes}, {\em {Parallel invaded
  cluster algorithm for the Ising model}}, 1998, cond-mat/9806127.

\end{thebibliography}

\end{multicols}

\end{document}